\documentclass[apjl]{emulateapj-rtx4}
\def\lesssim{\lower.5ex\hbox{$\; \buildrel < \over \sim \;$}}
\def\gtrsim{\lower.5ex\hbox{$\; \buildrel > \over \sim \;$}}
\def\UV{\mbox{\scriptsize UV}}

\shorttitle{Local absorption of GRB emission}
\shortauthors{Gilmore, R. and Ramirez-Ruiz, E.}

\begin{document}
\title{Local absorption of high-energy emission from gamma-ray bursts}

\author{R.~Gilmore\altaffilmark{1,2}, E.~Ramirez-Ruiz\altaffilmark{3}}

\altaffiltext{*}{Correspondence should be sent to R.~Gilmore (rgilmore@sissa.it)}
\altaffiltext{1}{SISSA, via Bonomea 265, 34136 Trieste, Italy}
\altaffiltext{2}{Department of Physics, University of California, Santa Cruz}
\altaffiltext{3}{Department of Astronomy, University of California, Santa Cruz}

\begin{abstract}
High-energy photons emitted from gamma-ray bursts (GRBs) are subject to pair-production interactions with lower energy photons, leading to an effective optical depth.  In this paper, we estimate the opacity resulting from photon fields located at various distances from long GRB sites:  that of the binary companion to the massive stellar progenitor, that of the star-forming molecular cloud containing the GRB, and the total photon field of the host galaxy.  The first two photon fields are found to be transparent for most reasonable sets of assumptions about these systems.  In the case of galactic radiation fields, we have performed several numerical simulations to calculate the expected opacities for 2 different line-of-sight geometries through the host galaxy, and include a full accounting of the infrared radiation produced by the absorption and re-radiation of starlight by dust.  The optical depth for GeV gamma-rays, due to direct starlight is less than unity for all host galaxies.  At higher energies, $>$10 TeV, a spectral cutoff can occur due to the rapidly increasing number of mid- to far-IR intra-galactic photons reradiated by dust.  Photons in the extragalactic background light therefore remain the only relevant  source of photon-photon opacity for ongoing GRB observations with Fermi LAT, and potential future detections with ground-based gamma-ray telescopes.  

\end{abstract}


\section{Introduction}
Recent observations by the Fermi satellite have confirmed that both long and short gamma-ray bursts are capable of emission at energies $>10$ GeV.  The rest-frame energies inferred for gamma rays from long-duration gamma-ray burst (GRBs) 080916C and 090902B \citep{abdo09c,abdo09a} and short GRB 090510 \citep{omodei09} are all in excess of $\sim$60 GeV.   These findings raise the question of whether emission might occur at still higher energies, and these gamma rays might be observable by Fermi, or by ground-based instruments such as MAGIC, VERITAS, and HESS.  Electron-positron pair-production from interactions between gamma rays and UV-IR photons \citep{gould&schreder67,stecker92,madau&phinney96,gilmoreUV} is an important effect at these energy scales, modifying the spectra of distant extragalactic sources and shrouding the highest-energy emission. This effect could thus be a powerful technique for understanding cosmological background radiation \citep{gilmoreGRB,fermiEBL}

The photon-photon optical depth per unit path length for a gamma ray of energy $E$ is 
\begin{eqnarray}
\lefteqn{\frac{d\tau(E)}{dl} =  \int^1_{-1}d(\cos\theta) \; \frac{(1-\cos\theta)}{2}} \nonumber \\ 
& & \times \int^{\infty}_{E'_{min}} dE'\; n(E')\;\sigma(E,E',\theta).
\label{eq:opdep}
\end{eqnarray}
where $n(E')$ is the local density of target photons at energy $E'$, and $\theta$ is the angle of interaction.  Here $\sigma(E,E',\theta)$ is the cross section for interaction \citep{madau&phinney96}, which peaks at twice the minimum energy for pair creation 
\begin{equation}
E'_{min}=\frac{2m_e^2c^4}{E (1-\cos\theta)}. 
\label{eq:mineng}
\end{equation}
\noindent At energies below 1 TeV, it is the optical, UV, and near IR photons that are responsible for attenuation.  Higher energy gamma rays have sufficient energy to produce electron positron pairs in interactions with longer-wavelength IR photons.  

In general, attenuation at a cosmological scale is expected to dominate the pair-production opacity, as the strength of the effect scales with path length through the intervening photon field.  Analyses of the interstellar radiation fields within the Milky Way find significant attenuation of high energy photons only at several tens of TeV, with negligible absorption at lower energies \citep{moskalenko06}.  There is strong evidence, however, that long-duration GRBs are associated with deaths of massive stars \citep{woosley&bloom06}, and therefore occur in star-forming regions of galaxies, and that the hosts of these objects often exhibit considerable star-formation rates \citep{bloom02,christensen04,fruchter06}, and therefore significant number densities of UV and optical photons.  The prevailing theory for these bursts is that they are associated with some subset of type 1c supernovae resulting from the deaths of short-lived stars \citep{woosley93,iwamoto98,macfadyen01}.

In this paper, we examine the optical depth for gamma-rays from a GRB produced by interactions with ``local'' photons within a host galaxy.  Loosely speaking, the opacity from pair-production is proportional to the product of path length and density of target photons at the relevant energy scale.  If the line-of-sight to the GRB passes through a region of intense flux from nearby stars, the resulting optical depth could conceivably be equal to or greater than that from intergalactic background photons.  Attenuation effects from photon fields on several different length scales could be pertinent.  In particular, we will discuss here absorption at sub-parsec scales from the photon field of a binary companion to the GRB progenitor, parsec-scale absorption within the star-forming region of the progenitor site, and finally opacity due to the kiloparsec-scale galactic disk of the host galaxy.
In \S \ref{approximations}, we present some simple estimates of the opacity arising on each of these scales.  In \S \ref{numcalcs}, we describe a numerical calculation performed to calculate the attenuation for star-forming galactic disks, followed by our conclusions.

\section{Gamma-ray attenuation at galactic and sub-galactic scales}
\label{approximations}
As a simple preliminary approximation, an estimate of $\tau_\gamma$ can be found for a given region by the product
\begin{equation}
\label{eq:tauest}
\tau_\gamma \approx n_{\UV} R \sigma,
\end{equation}
where $R$ is the typical size of the region of interest and therefore the approximate pathlength of the gamma ray through the photon field, $n_{\UV}$ is the number density of photons near energies corresponding to the peak cross section (maximum in $\sigma(E,E',\theta)$), and $\sigma \approx 0.1 \sigma_T$ is this approximated cross section.  Here $\sigma_T$ is the Thomson cross section.

\subsection{Binary companion}
Here we consider the attenuation of gamma rays by a high-mass binary companion of the GRB progenitor. One class of GRB creation models invokes a companion star at a few stellar radii to form a rapidly spinning helium core \citep{izzard04,podsiadlowski04,ramirez04,barkov10,yoon10}, a distance of $10^{10.5}$ to $10^{11.5}$ cm for a high-mass stars.  The effect of radiation from this companion on gamma-ray flux is limited by the radius $R_\gamma$ at which the high energy flux is emitted from the outflow of the progenitor. Most models place this emission at a distance $\gtrsim 10^{12}$ cm \citep{piran04,granot08,kumar09}, typically greater than the binary separation.  We have then for the opacity 

\begin{equation}
\tau_\gamma \propto \frac{ L_*}{4\pi} \int^{\infty}_{R_\gamma}  \frac{dR}{R^2} \; (1-\cos \theta),
\end{equation}

\noindent where  $L_*$ is the companion luminosity, and $\theta$ is the angle of photon-photon interactions from Equation (\ref{eq:opdep}).  If we consider the opening angle of the jet from relativistic beaming $\theta_0$, then the maximum $\gamma \gamma$ interaction angle possible is \[\theta \approx R_{\rm bin}/R_\gamma + \theta_0 \sim R_{\rm bin}/R_\gamma + \Gamma^{-1}.\]   The radius $R_\gamma$, for models in which the burst is triggered by the action of the companion, is expected to be larger than the progenitor--companion distance $R_{\rm bin}$.  Meanwhile the high bulk Lorentz factor, $\Gamma \gtrsim 100$, in most GRB models \citep{piran04} leads to a tightly beamed radiation cone.  This angular factor therefore greatly reduces the likelihood of interaction for GeV-scale GRB emission.

\begin{figure}[htbp]
\plotone{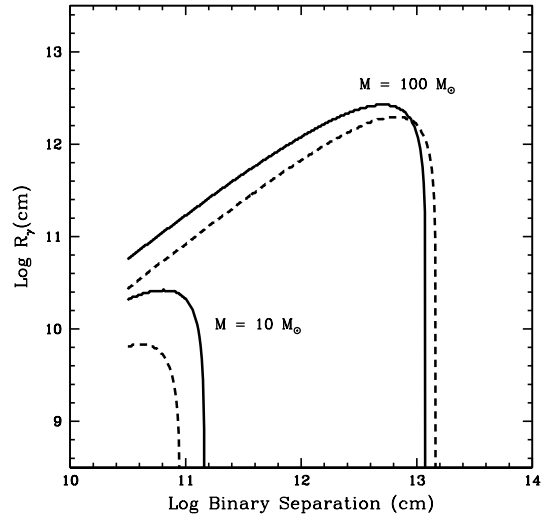}
\caption{Minimum gamma-ray production distance $R_\gamma$ from a close binary pair at a given separation for which $\tau_\gamma \lesssim 1$.  The solid and dotted lines are for gamma-ray energies of 300 and 100 GeV, respectively.  The lower and upper pairs of curves are for a low-metallicity main-sequence companion star of mass 10 M$_\odot$ and 100 M$_\odot$, see text for details.}
\label{fig:odbin}
\end{figure}

In Figure \ref{fig:odbin}, we show a summary of our results for gamma-ray opacity based on binary separation and emission radius, assuming that the direction of emission is perpendicular to the orbital plane.  Lines in this plot indicate the minimum gamma-ray production distance at which photon-photon opacity is less than unity, for a given separation distance between the progenitor and its companion.  We assume here a low-metallicity main-sequence companion at two different stellar masses, 10 and 100 M$_\odot$, which are modeled as blackbody sources.  From  \citet{tout96}, we assign these two cases luminosities of $10^{3.75}$ and $10^6$ L$_\odot$, and effective blackbody temperatures of $10^{4.5}$ and $10^{4.75}$ K.  These results suggest that while significant attenuation of gamma-rays can occur very close to the progenitor, it is unlikely to be dominant at the the radius at which gamma-ray are usually thought to be produced.

\subsection{Molecular cloud}
\label{mc}
For the case of attenuation by the photon field within a large molecular cloud, we can look to surveys of these complexes within the milky way (MW) and nearby galaxies to estimate the extent and magnitude of a typical photon field.  Clouds in the MW typically have sizes $R_{\rm mc}$ of $10^{0.5}$ to $10^{1.5}$ pc, and an H$_2$ gas mass as high as $\sim 10^6$ to $10^7$ M$_\odot$ \citep{solomon87,heyer01}, and other authors confirm similar results for M31 and the Large Magellanic Cloud \citep{rosolowsky07,fukui08}.  With the assumption that 5 to 10\% of this gas forms stars within a time comparable to the lifetime of high-mass stars, 20-30 Myrs \citep{krumholz06}, we can estimate the highest value likely for our estimate of $\tau_\gamma$ in Equation (\ref{eq:tauest}).

For a Salpeter IMF \citep{salpeter55}
\begin{equation}
\frac{dN}{dM} \propto M^{-2.35} \; ; \; \; 0.1\: \mbox{M}_\odot < M < 100\: \mbox{M}_\odot
\end{equation}
we have $\sim 0.005$ high-mass ($\geq 10$ M$_\odot$) stars per solar mass of gas converted into stars.  If we assume that these stars have a typical luminosity of $L_* =10^4$ L$_\odot$ which is emitted mostly in the UV and is responsible for most of this radiation at these wavelengths, then we estimate that the most massive molecular clouds could have $N_* = 5\: \mbox{x}\: 10^3$ massive stars producing $N_* \, L_* = 5\: \mbox{x}\: 10^7$ L$_\odot$ in optical-UV photons. The energy density in the cloud can be approximated as
\begin{equation}
\rho_{UV} = \frac{N_* \, L_*}{R_{\rm mc}^3} \left(\frac{R_{\rm mc}}{c}\right),
\end{equation}
about 6.7 x $10^{-9}$ erg cm$^{-3}$ under our assumptions.  If we assume that this energy exists in the form of UV-optical (i.e. 3 eV) photons, in a cloud of size $R_{\rm mc} = 10$ pc, then we have an photon number density of 1.4 x $10^3$ cm$^{-3}$.  Applying Equation (\ref{eq:tauest}) to find the opacity over a path through the cloud, we estimate \[\tau_\gamma \sim 3 \; \mbox{x} \; 10^{-3}\] from UV photons in the most massive stellar birth clouds.  We have not accounted for dust extinction in these systems, which can be substantial and would reduce the UV flux and $\tau_\gamma$.

\subsection{Galactic Disk}
\label{galdisk}
While it is generally accepted that long-duration GRBs are associated with core-collapse supernovae \citep{woosley&bloom06,gehrels09}, the relation between GRB host galaxies and the population of star-forming galaxies appears to be complex, and recent studies seem to dismiss the idea that GRB rates follow star formation in an unbiased manner \citep{ramirez02,cen&fang07,guetta&piran07}.  Surveys of host galaxies can provide a guide for us to choose general parameters in this project.  A survey of GRB host galaxies at a variety of redshifts by \citet{wainwright07} found that hosts can typically be described by an exponential profile with a median scale radius of 1.7 kpc; none followed the de Vaucouleurs profile associated locally with elliptical galaxies.  

\citet{castroceron08} analyzed 30 GRB hosts, and found typical star-formation rates of 0.01 to 10 M$_\odot/$yr from an assumed unobscured UV, though the inclusion of dust could raise these values considerably.  This paper also mentions that some galaxies in their sample with more poorly constrained SEDs could have star formation rates as high as a few hundred M$_\odot/$yr.    A study of 46 hosts by \citet{savaglio09} found star-formation rates of 0.01-36 M$_\odot$/yr, with a mean of 2.5, after correction for dust extinction.  Stellar masses in these samples ranged from $10^7$ to $10^{11}$ M$_\odot$ in the former, and $10^{8.5}$ to $10^{11.1}$ M$_\odot$ in the latter.  However,  

\citet{perley09} has found that even galaxy hosts for obscured ``dark'' GRBs typically have normal optical colors overall, suggesting that the dust is unevenly distributed or local to the GRB.  In their survey, \citet{castroceron08} estimate that $\sim$25\% of GRB hosts have dust extinction A$_{\rm V} \geq 1$.  \citet{savaglio09} found an average dust extinction of $<$$A_V$$>\sim 0.5$ in a subset of 10 galaxies in their sample; 2 of these galaxies had $A_V >1$.

Following our method of the last two sections, we can perform an order-of-magnitude estimate of $\tau_\gamma$ for a rapidly star-forming galaxy.  For a given star-formation rate, we can estimate the UV flux using the approximation \citep{madau98}
\begin{equation}
L_{\mbox{UV}} \approx 8 \: \mbox{x} \: 10^{27} \; \frac{\mbox{SFR}}{\mbox{M$_\odot /$ yr}} \; \mbox{erg}\; \mbox{s}^{-1}\: \mbox{Hz}^{-1}.
\end{equation} 
\noindent For a star-formation rate of 100 M$_\odot/$yr, and a galaxy size of 1 kpc, we estimate a total UV output of 2 x $10^{45}$ erg/s, and  consequently a photon energy density similar to that of the molecular cloud in the previous section.  From Equation (\ref{eq:tauest}), we estimate a gamma-ray opacity \[\tau_\gamma \sim 0.3 \] if this energy is emitted as 3 eV photons.

These finding suggest that the emission from a star-forming galaxy, acting over a path length of a kpc or more, could be a barrier to GeV gamma-rays from a GRB within a galaxy.  However, this is only an order-of-magnitude estimate of $\tau_\gamma$, with a high star-formation rate and no account of the absorption and re-emission of starlight by dust.   Higher star-formation rates than we assume here have been seen in GRB-selected submillimeter galaxies \citep{michalowski08}, and in extreme cases galaxies of this type can have rates exceeding 1000 M$_\odot/$yr \citep{daddi05}, albeit with heavy dust obscuration.  To better understand the circumstances under which the photon flux from stars and dust in a galaxy might lead to significant attenuation of gamma rays on a kiloparsec scale, we have performed a set of numerical calculations which are described in the next section.

\section{Numerical calculations of opacity from stellar disks}
\label{numcalcs}
\subsection{Model}
In order to accurately estimate the contribution to opacity arising from a compact, rapidly star-forming galaxy, we have performed a simple simulation of a galactic disk, and calculated numerically the resulting UV-IR photon field and from this the gamma-ray optical depth along a given line-of-sight.  The stellar spectra are calculated using the population synthesis code of \citet{bruzual&charlot03}.  We assume that the stellar population in our galaxy is formed at a constant rate over a characteristic timescale $T_{\rm sf}$.  Stars are then distributed in a galactic disk of exponential scale length $R_e$ and maximum radius $3R_e$, and with thickness $0.3R_e$.  Note that the half-light radius for a exponential profile is $1.68 R_e$.

Our calculation includes a simple model for dust absorption and re-emission.  Starlight is absorbed using the model of \citet{charlot&fall00}, in which starlight is absorbed by 2 components: an interstellar medium (ISM) component affecting all stars, and a molecular cloud component affecting only stars younger than $10^7$ yrs.  We assume the standard ratio for these opacities, in which the ISM is responsible for 30\% of the opacity for young populations, and the rest arises from absorption within molecular clouds.  The ISM dust component is assumed to have a homogeneous distribution in the same disk as the stellar population.  All energy absorbed by dust is re-emitted in the IR.  We calculate the emission at wavelengths above 4 $\mu$m using the templates of \citet{rieke09}, which are based on {\it Spitzer} observations of local galaxies.  We consider face-on dust extinction factors of 0.1, 1.0, and 3.0 in the V-band attenuation $\tau_{\rm V}$, broadly corresponding to very low, moderate, and large amounts of reddening.  From this normalization, the extinction at other wavelengths is determined by multiplying by a power law in wavelength with index -0.7.  In galaxies with $\tau_{\rm V} = 3.0$, greater than 90\% of starlight from stellar populations younger than 500 Myrs is absorbed by dust and re-emitted in the IR, so this corresponds to a (U)LIRG-like mode with highly-obscured star-formation.   All of our galaxy spectra are created using a metallicity of [Fe/H] = -1.65.  This is somewhat less than the typical metallicity of 1/6 solar that was seen in the sample of \citet{savaglio09}; we do not expect this to notably impact our final results.

\begin{figure}[htbp]
\plotone{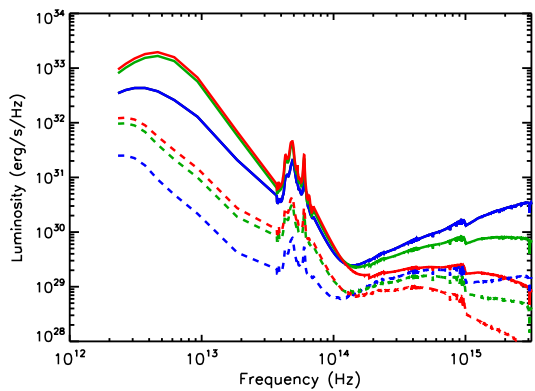}
\caption{The emission spectra for some of the galaxies in our model.  The solid lines show are for a stellar population with $T_{\rm sf} = 30$ Myr, while the dashed lines are for $T_{sf} = 1$ Gyr.  Line colors indicate different levels of dust attenuation, as described in the text.  Blue, green and red correspond to extinction factors of 0.1, 1.0, and 3.0 in $\tau_{\rm V}$ (top to bottom, respectively, at $10^{15}$ Hz).   Stellar masses have been normalized to $10^{10} M_{\odot}$.}
\label{fig:galspect}
\end{figure}

\begin{figure}[htbp]
\plotone{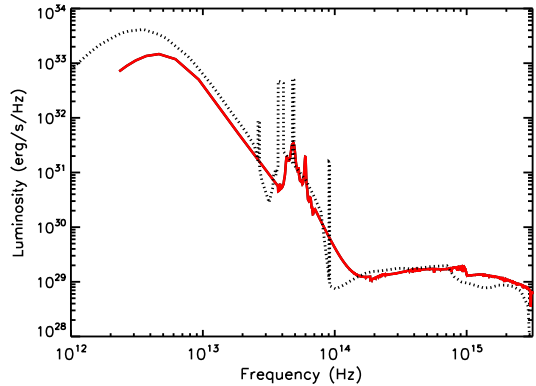}
\caption{Comparison between one of our model spectra and that of the fit to GRB 000418 shown in \citet{michalowski08} (black dotted line).  Our spectrum (solid red line), is for a total dust extinction of $\tau_{\rm V}$ = 3.0, and has been normalized to match the inferred star-formation rate for this host of 288 M$_\odot/$yr.}
\label{fig:galspect_comp}
\end{figure}

Figure \ref{fig:galspect} shows the sample spectra for galaxies of two different star-formation timescales ($T_{\rm sf}$) and three values for face-on dust extinction ($\tau_{\rm V} = 0.1$, 1.0, and 3.0), normalized to the same mass.  For our calculation, we have considered timescales over the range 10 Myr $\leq T_{\rm sf} \leq$ 10 Gyr.  In Figure \ref{fig:galspect_comp}, we show a comparison with the SED for a starbursting GRB host, GRB000418, as presented in \citet{michalowski08}.  This source is representative of the type of GRB hosts that have the highest inferred star-formation rate and are therefore expected to have the largest opacity to gamma rays. 

We do not include a luminosity contribution from AGN in our calculation.  However, our templates for dust re-emission are based on observations of local galaxies, and therefore some degree of AGN contamination in the IR unavoidably exists.  Some degree of evolution in the typical IR properties of galaxies likely occurs between local redshifts and those near the peak of the star-formation era (z$\sim$2); \citet{rieke09} estimates that these are not larger than a factor of two.  
In our model we assume a scaling law for the galactic disk radius with stellar mass of 
\begin{equation}
R_e \propto M_\ast^{1/3}
\label{eq:galmasssize}
\end{equation}
with normalization of 500 pc at $10^9$ M$_\odot$; this slope is based upon a rough fit to the half-light radius results from a survey of star-forming galaxies in \citet{forsterschreiber09}, and is consistent with the more compact galaxies seen in the sample presented in \citet{trujillo04}.  This particular scaling produces galaxies that are more compact than 18 of 20 GRB hosts with $r_{80}$ ($\approx 3R_e$) fits in the study of \citet{svensson10}. Since the disk height is also proportional to $R_e$ in our model, it should be pointed out that for a given set of stellar and dust properties the photon flux density scales as $R_e^{-2}$, and therefore gamma-ray opacity is proportional to $R_e^{-1}$, after accounting for the increase in path length.  The UV, optical, and near-IR spectral properties of a unit mass of stars are defined in our model by the assumed dust attenuation factors and star-formation timescale, therefore for gamma-ray energies affected solely by these wavelengths ($\lesssim 500$ GeV) the gamma-ray opacity scales as
\begin{equation}
\tau_\gamma \propto \mbox{M}_\odot^{2/3}
\end{equation}
when these factors are held fixed.

\subsection{Results}
The primary ingredients of our model are the star-formation timescale, dust attenuation factor, and galaxy size and stellar mass which are conjoined in Equation (\ref{eq:galmasssize}), as discussed in the previous section.  The other important considerations are the location of the GRB within the disk and the geometry of the line-of-sight (LOS) path from the GRB to the observer.  We fix the location of the GRB at a radius $0.75 R_e$ (25 \% of the simulated radius $3 R_e$) and centered within the disk height.  We consider two possibilities here for the LOS:  a ``skimming'' mode in which the path is toward the galactic center at an angle of incidence of 10 degrees to the disk, and a geometry in which the path is face-on relative to the disk, with angle of incidence 90 degrees.  The former gives a nearly maximal amount of attenuation for a given galaxy, while the second gives a closer to average amount of absorption.

\begin{figure}[htbp]
\plotone{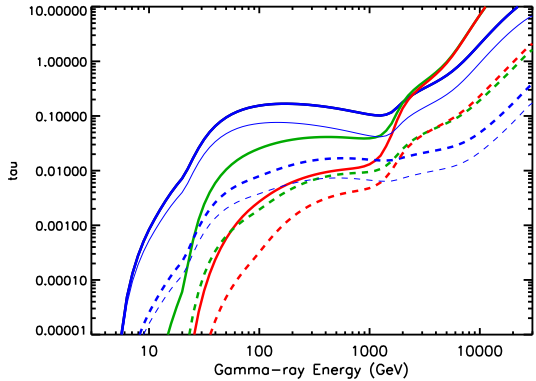}
\caption{Optical depth for gamma rays created by the photon field in and around a galaxy of given stellar properties, for a line-of-sight along the galactic disk.  Line types and colors are the same as in Figure \ref{fig:galspect}; solid and dashed lines are for star-formation timescales of 30 Myr and 1Gyr, and blue, green and red are for dust extinction factors of 0.1, 1.0, and 3.0 in $\tau_{\rm V}$ (top to bottom, respectively, at 100 GeV).  Thin lines for the blue curves show how the attenuation changes when the path is instead perpendicular to the galactic plane.  Results here are again for a galaxy of stellar mass $10^{10}$ M$_\odot$.}
\label{fig:opdep}
\end{figure}

We have calculated opacities for gamma rays from energies of 1 GeV to 100 TeV.  In Figure \ref{fig:opdep}, we show the opacity $\tau_\gamma$ for gamma rays at a particular energy for several sets of galaxy properties.   Changing from the skimming LOS to one that is perpendicular to the galactic disk is found to reduce $\tau_\gamma$ by a factor of 2 -- 3. 

\begin{figure}[htbp]
\plotone{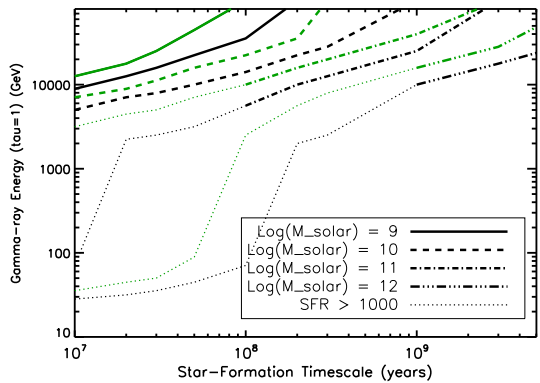}
\caption{The minimum gamma-ray energy for which $\tau_\gamma \geq 1$, as a function of star-formation timescale in the host galaxy.   In this plot we show results for a low level of dust extinction, $\tau_V = 0.1$  Line types indicate various galaxy stellar masses, and we have assumed a characteristic radial scale length for the disk from Equation \ref{eq:galmasssize}.  Line color indicate results for the different line-of-sight geometries, black is for a direction skimming along the disk, and green is for a direction perpendicular to the disk, as discussed in the text.  The thin dotted lines denote star-formation timescales that imply recent star-formation rates in excess of 1000 M$_\odot/$yr for the given stellar mass. }
\label{fig:opthick1}
\end{figure}

\begin{figure}[htbp]
\plotone{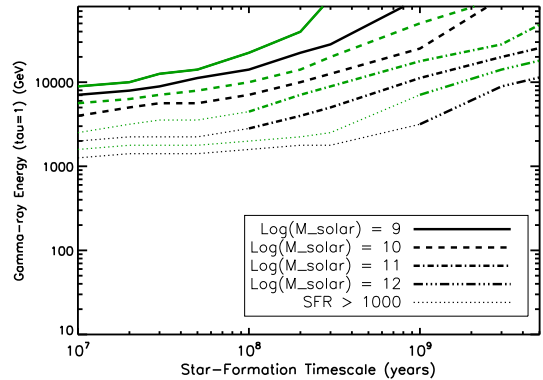}
\caption{As in Figure \ref{fig:opthick1}, but for total dust extinction coefficient of 1.0.}
\label{fig:opthick2}
\end{figure}

The results in Figure \ref{fig:galspect} for a galaxy of mass $10^{10}$ M$_\odot$ indicate that attenuation at energies $<$ TeV is significantly less than $\tau_\gamma=1$ even for the optimal ``skimming'' geometry and a fast stellar buildup time of 30 Myrs.  As mentioned in the previous section, the attenuation is expected to be approximately proportional to $M_\ast^{2/3}$, if we assume that galaxy volume scales with mass.   The only significant attenuation in these cases is due to the population of mid- and far-IR photons created by thermal and PAH emission; these being much more numerous than the UV, optical, and near-IR photons of direct starlight.

In Figures \ref{fig:opthick1} and \ref{fig:opthick2} we examine the relationship between galaxy mass, star-formation timescale (or star-formation rate) and the minimum gamma-ray energy for which $\tau_\gamma \geq 1$; these two plots are respectively for cases of low and moderate dust extinction ($\tau_V=$ 0.1 and 1.0).  Results for the largest dust extinction, $\tau_V=3.0$, are not substantially different from the 1.0 case and are not shown.  Displaying our results in this way can tell us the approximate energy scale at which photon-photon opacity becomes an important effect on the high-energy spectrum of radiation escaping from the vicinity of the galaxy.  In general, the opacity only becomes significant at multi-TeV energies, where gamma-rays have sufficient energy to interact with the re-emitted IR photons.  As seen in Figure \ref{fig:opdep}, the opacity curve is approximately flat between 100 GeV and 1 TeV.  Opacity at these energies is generally less than 1,  and we only find optically-thick galaxies in cases of minimal dust extinction, high mass, and short star-formation timescale, with star-formation rate $\gtrsim 10^4$ M$_\odot/$yr.

\section{Discussion} 

Our results suggest that it is unlikely the radiation field within a single star-forming region could impact the high-energy (GeV and TeV) gamma-ray emission from a GRB, although our understanding of these regions is limited to nearby galaxies.  Significant attenuation due to the emission of a massive binary companion is possible only if the radius of high-energy photon emission is  less than predicted by most radiation models.

Large, rapidly star-forming galaxies can form an optically-thick barrier only at multi-TeV energies, where attenuation with mid-IR photons can occur, regardless of line-of-sight geometry or dust extinction.  However, photons at these energies would already be strongly attenuated by extragalactic background light for all but the closest GRBs (z \lesssim{0.05}).  Creating enough photons in the UV, optical and near IR to significantly impact transmission of GeV photons is only possible with star-formation rates well in excess of 1000 M$_\odot$/yr, even with a minimal amount of dust attenuation.  This requirement could be lowered if galaxies are much more compact than we have assumed, though the galaxy sizes used here are already more compact than most of those in the sample of \citet{trujillo04}, and these authors mention that the densest objects at high redshift tend to be elliptical galaxies with much lower photon densities.  

Clumpy or irregular star formation is one possibility that we have not considered in our numerical calculations, which assume a smooth exponential disk.   The total amount of attenuation through the host galaxy is proportional to the photon density over the line-of-sight path.  Therefore the total $\tau_\gamma$ for the galaxy would only be increased significantly from our predictions if a large fraction of the total stars in the galaxy were placed in a much more compact distribution than we have assumed with our exponential profiles.  This region would have to contain a much higher density of young stars than the star-forming regions seen in the MW and LMC, based on the order-of-magnitude analysis of \S \ref{mc}.  Merely rearranging star formation into clumpy regions throughout the galaxy would not, in general, affect the total photon field or $\tau_\gamma$, unless those regions were were positioned near the line-of-sight path in a stochastic irregularity.  In conclusion, both long and short GRB observations can be safely used as powerful tools  for understanding the extragalactic background light .

\acknowledgments
We thank J.~Granot, X.~Prochaska, and J.~Primack for useful discussions, M.~ Michalowski and D.~Watson for providing us with their spectral fits to GRB host galaxies, and the anonymous referee for comments which helped improve this paper.  R.G. acknowledges support from a Fermi Guest Investigator Grant during part of this research. E.R-R acknowledges NSF grant AST-0847563, and thanks the Packard Foundation for support.

\end{document}